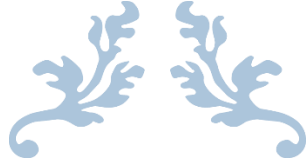

# A network Based method to predict cancer causal genes in GR Network.

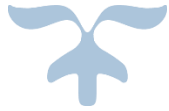

Prepared by :

Mostafa Akhavansafar
School of Systems and Industrial Engineering, Tarbiat Modares University (TMU), Tehran, Iran.
Email: m.akhavansafar [at] modares.ac.ir

Babak Teimourpour
**Corresponder:** Assistant Professor, Room No. 210, 2nd Floor, Information Technology Engineering group, School of Systems and Industrial Engineering, Tarbiat Modares University (TMU) Chamran/Al-e-Ahmad Highways Intersection, Tehran, P.O. Box 14115-111, Iran.
E-mail address: b.teimourpour [at] modares.ac.ir
Tel: +98(21) 8288 3987, Fax: +98(21)8288 4323

March 09, 2020

TABLE OF CONTENTS

LIST OF TABLES

LIST OF FIGURES

# SECTION I

# SECTION II

# Abstract

One of the important issues in oncology is finding the genes that perturbation the cell functionality, and result in cancer propagation. The genes, namely driver genes, when they mutate in expression, result in cancer through activation of the mutated proteins. So, many methods have been introduced to predict this group of genes. These are mostly computational methods based on the number of mutations of each gene. Recently, some network-based methods have been proposed to predict Cancer Driver Genes (CDGs). In this study, we use a network-based approach and relative importance of each gene in the propagation and absorption of genes anomalies in the network to recognize CDGs. The experimental results are compared with 19 previous methods that show our proposed algorithm is better than the others in terms of accuracy, precision, and the number of recognized CDGs.

KatzDriver: A network Based method to predict cancer causal genes in GRN

Section I

Introduction to the Study

## Introduction

Cancer is one of the serious challenges, and the researchers perform many efforts to cope with and treat it. With technology improvement and advanced tools of sequencing, a high amount of data has been obtained, and the disease treatment has been changed from a molecular approach to a network-based approach. In this approach, the disease is considered as a molecular system, and the goal is to find the anomalies that make perturbation in the total functionality of the system. One of the important issues in oncology is the detection of the cancer-causing genes [1]. In a cell, different components communicate with each other and create a biological system [2]. One of these systems is the gene regulatory network. In these networks, the effect of genes on each other leads to the change of the gene expression rate and protein production in the cell [3]. The change of a gene expression rate makes changes in the expression rate of other genes. The expression rate of these genes and the changes should not get the cell functionality out of normal situations. A mutation in an expression rate of one or some of the genes and a problem in the cell's regulatory function, result in cancer. All the performed mutations don't result in cancer, and some of the mutations named driver result in cancer. The driver mutations lead to activation of the mutated proteins that are involving in signaling paths. So the mutated proteins make the tumorigenesis induction and its survival. The genes that their mutations lead to cancer progression are Cancer Driver Genes (CDG). CDGs based on their role in disease is classified into Tumor suppressor genes or oncogenes. Different methods have been proposed to detect CDGs. Most of the proposed approaches are based on computational methods and use mutation concept such as ActiveDriver, OncodriverCLUST , e-Driver, Oncodrive-fm, Simon, Dendrix, CoMDP, MDPFinder,  MutSigCV, iPAC, MSEA, DrGap, r-Driver , and ExInAtor. In some of the previous techniques like DriverNet , NetBox , DawnRank, MeMo, and SCS in addition to mutation, network concept is used. Some other methods like iMaxDriver-N, iMaxdriver-w , and GHTS  use only the network for CDG recognition. In computational methods, based on analysis of the number of copies and gene expression data, a sequence of statistical tests is performed in a list of genes systematically to extract the cancer cells list. It means the driver genes are recognized through repetitive mutations in different tumor samples. The main challenge of these methods is the separation of cancer and non-cancer mutations. The problem is solved to some extent using the network and system approaches. For example, iMaxDriver methods use the maximization of the propagation approach in the network to predict cancer genes. GHTS detects the driver genes by modifying the web pages' search algorithms based on the biological data, gene expression, and calculation of each gene penetration rate. In MSEA, network features are used along with genomic and biologic features. DawnRank uses data interaction information to improve the performance of gene networks in addition to expression and mutation data. In Memo, CDGs recognition is performed based on the calculation of the correlation between genes and forming a network among the genes. DriverNet recognizes the driver genes by the relation between the genomic patterns and transcriptional data and analysis of the penetration network. So, the literature review shows that the previously proposed methods go from computational methods to the semi-networking and network-based methods, because of the performance improvement and using intrinsic features of molecular networks that are less considered in computational and mutation-based methods.

Also, the mentioned computational and networking methods have some disadvantages. In most of the computational methods, the amount of false-positive is high, and precision and *F-measure* are low. Moreover, their performance in terms of the number of recognized CDG is not desirable. The network-based methods, also, have some problems. For example, iMaxDriver-N and iMaxDriver-W have very high computational complexity, also weighting to the interactions is random, which is not a precise weighting method based on gene expression changes. Moreover, the method only recognizes TFs with the CDG role, while some of the CDGs include mRNA. GHTS is another network-based method that solves the computational complexity and random weighting of iMaxDriver methods. It has better performance than other methods in terms of the number of detected CDGs and F-measure. But, it considers only the TFs with driver role without considering the part of mRNAs that are a component of CDG.

In this research, a network-based algorithm called KatzDriver is proposed to recognize CDGs that can detect cancer-cause mRNAs along with cancer-cause TFs without the need for information on the biologic path and mutation data. Moreover, it eliminates the disadvantages of previous network-based methods and improves recognition performance.

## Study method

### Relative influence of a node in network

In graph theory, there are different methods to calculate the effect of a node in a network. One of the methods is using the Katz measure to measure the amount of relative effect of a node in a social network [4]. Katz calculates the relative effect of a node in the network by measuring all of the walks between two nodes. While all of the network measures only consider the shortest path between two nodes, it can better show the relative importance of the nodes. So, it is similar to the PageRank algorithm of Google. Katz calculates the relative effect of a node in the network by measuring the number of direct neighbors and other nodes of the network that is connected to them through its direct neighbors. It is generalized centrality of eigenvector [5].

The relative effect of each node based on Katz centrality for node i is calculated through equation (1).

$$Katz_i = \alpha \sum_{j=0}^{n} A_{ij}.Katz_j + \beta \qquad (1)$$

Where A is the adjacency matrix. $\beta$ is used to prevent zero centrality and it is the weight that is assigned to direct neighbors. $\alpha$ is the damping factor for the connections with distant neighbors that should be lower than the reverse of the highest Eigenvalue of the adjacency matrix to correct calculation of Katz centrality ($\alpha < \frac{1}{\lambda_{max}}$). The difference between direct and indirect neighbors is determined using this parameter. In directed graphs, the centrality for a node is converted to two centralities of in-edges and out-edges. It shows the important of a node in the network in terms of influence power on the nodes and reception power from other nodes. It is defined by equations (2) and (3).

$$Katz_i^{out-edges} = \alpha \sum_{j=0}^{n} A_{ij}.Katz_j^{out-edges} + \beta \qquad (2)$$

$$\boldsymbol{C}_{Katz}^{out-edges} = \alpha A.\boldsymbol{C}_{Katz}^{out-edges} + \boldsymbol{\beta}$$

$$\boldsymbol{C}_{Katz}^{out-edges} - \alpha A.\boldsymbol{C}_{Katz}^{out-edges} = \boldsymbol{\beta}$$

$$\boldsymbol{C}_{Katz}^{out-edges}(I - \alpha A) = \boldsymbol{\beta}$$

$$\boldsymbol{C}_{Katz}^{out-edges} = \boldsymbol{\beta}.(I - \alpha A)^{-1}.\mathbf{1}$$

$$Katz_i^{in-edges} = \alpha \sum_{j=0}^{n} A_{ji}.Katz_j^{in-edges} + \beta \qquad (3)$$

$$\boldsymbol{C}_{Katz}^{in-edges} = \alpha A^T.\boldsymbol{C}_{Katz}^{in-edges} + \boldsymbol{\beta}$$

$$\boldsymbol{C}_{Katz}^{in-edges} - \alpha A^T.\boldsymbol{C}_{Katz}^{in-edges} = \boldsymbol{\beta}$$

$$\boldsymbol{C}_{Katz}^{in-edges}(I - \alpha A^T) = \boldsymbol{\beta}$$

$$\boldsymbol{C}_{Katz}^{in-edges} = \boldsymbol{\beta}.(I - \alpha A^T)^{-1}.\mathbf{1}$$

$$\boldsymbol{C}_{Katz}^{in-edges} = \boldsymbol{\beta}.(I - \alpha A^T)^{-1}.\mathbf{1}$$

## KatzDriver Algorithm to CDGs Detection

The studied network in this research is a transcription regulatory network. Such biological networks are obtained from the gene expression data. The nodes of the network are genes (including TFs and mRNAs), and the edges show the regulatory interactions between the nodes. There are two kinds of modules in these networks: gene modules and prescription modules. There are some genes in gene modules that all of them are regulated by one factor. In the prescription modules, there are some factors that all of them regulate common genes [3]. The regulatory interactions include TF-TFs and TF-mRNA.

The driver genes consist of both prescription factors and mRNAs. According to the network structure, TFs mostly have influencial property, and mRNA genes mostly have receptional property, and both of them have a role in cancer propagation. In a regulatory gene network, the genes may change the expression of each other by the effect on each other. If these changes exceed the normal bound, it is mutated. If the mutation is performed in the driver genes, it leads to cancer. So, we can consider the genes with the highest relative impact as the driver genes. Because if they mutate, they will have the most impact on the perturbation in the cell regulatory functionality, anomaly propagation, and cancer occurrence. Hence, the genes can be ranked by calculation the amount of their influence and reception power, and introduced the genes with highest scores as the cancer driver genes. We get two scores for each gene by modifying the Katz algorithm using the gene expression data and gene interactions weight. One is the relative effect of each gene in the gene regulatory network based on incoming interactions to it, and another one is the relative effect of each gene based on outgoing interactions from the gene. We called the new algorithm as KatzDriver.

## Network Construction

Gene regulatory interactions and expression data are needed to construct the network and applying the algorithm. For an accurate investigation of the results and show improvement of our method rather than GHTS, TRRUSTv2 database [6] is used to download the interactions list, and the GEO database is used to get gene expression data related to three cancers of colon, lung, and breast(Table 1). At first, weight is considered for each node using gene expression data using the Foldchange concept. Foldchange is a scale to determine the number of quantitative changes between the primary and secondary values of a variable, and it is considered as the rate between tow quantities. Foldchange means multifold. In other words, the considered gene expression in the first group samples (e.g., cancer samples) rather than the second group samples (e.g., normal samples) is multiplied or reduced several times [7]. If A and B are two values, and we want to calculate Foldchang for B based on the value of A, we should perform B/A. Foldchange is mostly performed when analysis of various measures of a biological system at different times. We extracted data related to gene expression of five patients in normal and cancer tissues for three cancers of breast, lung, and colon. Then we calculated the value of Foldchange for each gene by equation (4).

$$Fold\ change(\ gene_i) = \frac{\sum_{i=1}^{5} Cexp_{gene_i}}{\sum_{i=1}^{5} Nexp_{gene_i}} \tag{4}$$

Then using the resulted values for each gene, we construct $\vec{v}$ vector include foldchange values of genes. The weight of regulatory transactions can be calculated by minimizing the equation (5) that is proposed in GHTS.

$$\ell = Min||L^T.v - v|| \qquad \vec{v}: vector\ of\ foldchange\ value \tag{5}$$

Moreover, the initial score of the relative effect of each node is calculated based on the sum of the weights of outgoing edges and the sum of the incoming edges, and the value of Foldchange is calculated by (6).

$$instart_{gene_i} = |Fold\ change(\ gene_i) \times (\sum_{j\in out_{edges}} w_{ij} - \sum_{j\in in_{edges}} w_{ji}| \tag{6}$$

The relative value of the effect of each gene in the gene regulatory network computed through equations (7) and (8).

$$KTZd_{gene_i}^{out-edges} = \alpha \sum_{j=0}^{n} w_{ij}.KTZd_{gene_j}^{out-edges} + \beta \qquad (7)$$

$$KTZd_{gene_i}^{in-edges} = \alpha \sum_{j=0}^{n} w_{ji}.KTZd_{gene_j}^{in-edges} + \beta \qquad (8)$$

The considered values of $\alpha$ and $\beta$ are 0.1 and 1, respectively. The values of incoming and outgoing relative importance are combined as equation (9), to obtain the final relative effect of each gene in the network.

$$KTZd(gene_i) = \lambda \times KTZd_{gene_i}^{out-edges} + (1-\lambda) \times KTZd_{gene_i}^{in-edges} \qquad (9)$$

Where $0 < \lambda < 1$ and it is used to regulate the amount of the two scores effects on the final score. If $\lambda = 0$, the resulted scores show only the amount of each gene's power for information propagation and based on the network structure, it recognizes the TF-type cancer driver genes. If $\lambda = 1$, the resulted scores show only the amount of each gene's power for information reception and based on the network structure, it recognizes the mRNA-type cancer driver genes. But, when the calculation of only outgoing or incoming power of each gene, the system has not its best performance. So, in this algorithm, $\lambda$ is used as a trade-off for weighting between propagation and absorption powers of the genes anomalies. We expect that different values of $\lambda$ are effective on the outcome performance. To select the best value of $\lambda$ we used the area under the curve of ROC for each cancer with the values $\lambda = 0, \lambda = 0.1, \lambda = 0.2, \lambda = 0.3, \lambda = 0.4, \lambda = 0.5, \lambda = 0.6, \lambda = 0.7, \lambda = 0.8, \lambda = 0.9, and \lambda = 1.0$. Different values of AUC for breast, colon, and lung cancers are shown in figure 1. The higher level of the curve shows the model has a better distinction between cancer and non-cancer genes. The results show that in all the three mentioned cancer, when $\lambda$ is zero, AUC has its lowest value. For $\lambda = 1$ its performance is in the middle rank, that is when only the ability of information propagation and reception of a gene is considered, the model performance is not good to detect the driver genes. By increment of $\lambda$, AUC increases and it reaches its highest level, when $\lambda = 0.9$. This feature is for all the three mentioned kinds of cancer and $\lambda = 0.9$ is considered. In the next section, we show that the model has the best f-measure and the most recognition of CDGs, by considering $\lambda = 0.9$.

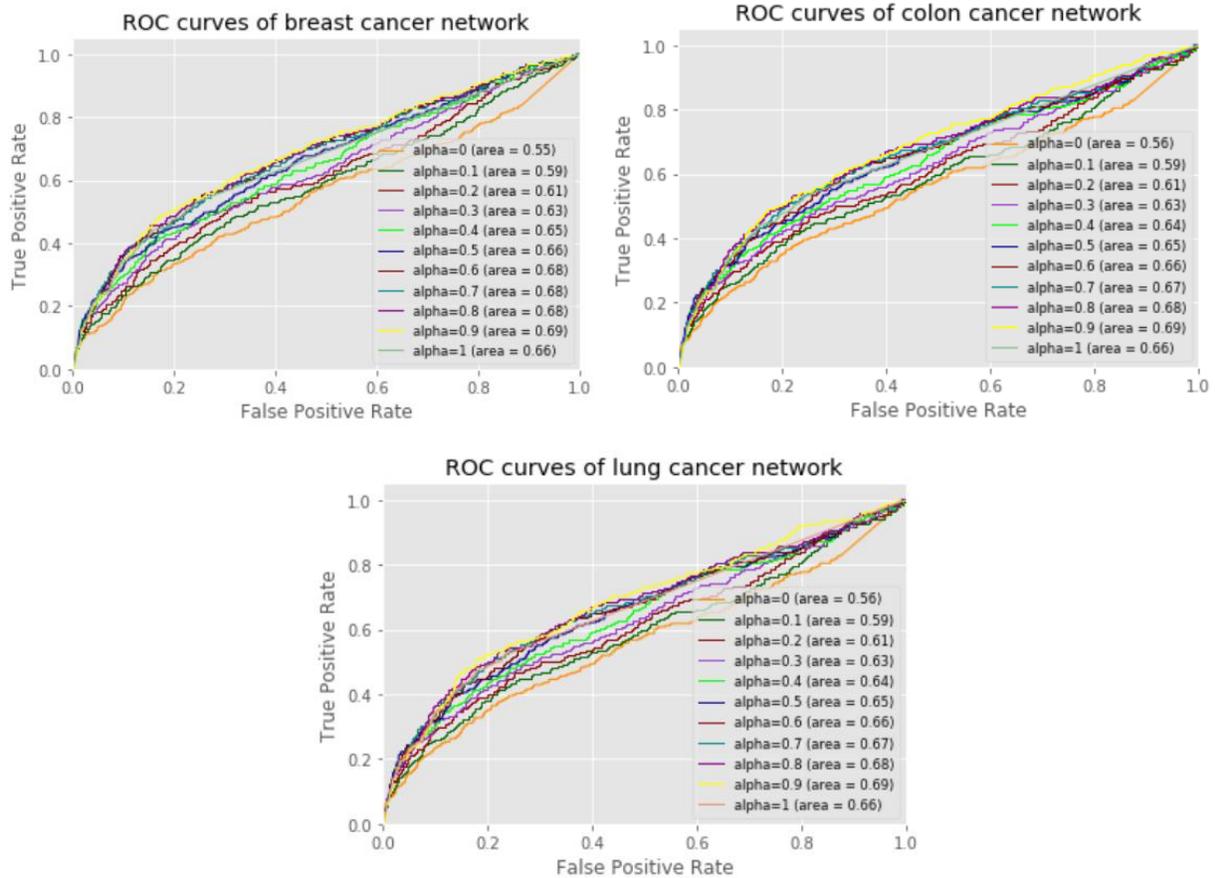

**Figure 1:** ROC curves of KatzDriver with different $\lambda$.

## 2-3 Data Processing and Evaluation Method

Utilized data in this research include regulatory interactions to create gene regulatory network and gene expression values to create biological value to the nodes and the edges. The regulatory interactions list and data of gene expression are extracted from TRUUST V2 and GEO [8] databases, respectively, for breast, lung, and colon cancers. After preparation of the gene expression data and filtering extra interactions of the list, regulatory interactions of the considered network for the three mentioned cancer are constructed. Based on the mentioned methods in section 1-2-2, the nodes and the edges are weighted. Moreover, the Cancer Gene Census (CGC) list [9] is used to compare the results consisting of driver genes for the three types of cancer. In this dataset there are 572, 566 and 572 CDGs for breast, Lung and colon cancer respectively. Moreover, the results are compared with the driver genes of MSKCC , the driver genes introduced by Vogelstein et al., namely Mut-driver and the driver genes introduced by Kumar et al., namely HiConf, to verify the results. In MSKCC There are 197, 397 and 423 CDGs for breast, Lung and colon cancer respectively. Mut-driver consist of 125 cancer driver genes. Also, HiConf consist of 99 CDGs. The set of driver genes of MSKCC is downloaded from the cBioPortal database (https://www.cbioportal.org), and the driver genes of Mut-driver and HiConf are obtained from their related papers.

F-measure, Recall, and Precision are used to evaluate the accuracy and precision of the proposed method.

$$Recall = \frac{TP}{TP + FN} \qquad\qquad (10)$$

$$Precission = \frac{TP}{TP + FP} \qquad\qquad (11)$$

$$F - measure = 2 \times \frac{Precision \times Recall}{Precision + Recall} \qquad\qquad (12)$$

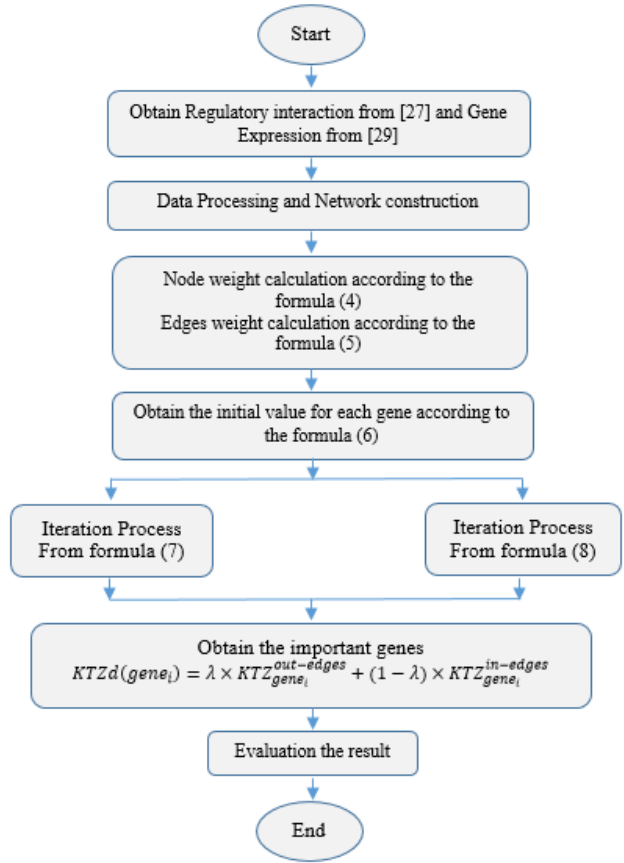

Figure 2: flowchart of KatzDriver algorithm

## Section II
## Findings, Conclusions, and Recommendations

### Introduction

This study was designed a network based approach, KatzDriver, to detect the cancer causal genes in gene regulatory network in human. We compared result of our method with 19 other computational and networking methods.This section includes the Findings, Conclusions, and Recommendations.

## Findings

Python programming language has been used for the implementation of the algorithm. Figure 2 shows the flowchart of the algorithm. By execution of the KatzDriver algorithm in each of the three cancer networks, an importance rank has been obtained for each gene that is a combination of its ability of propagation and information reception. Then ranked genes were sorted in descending order. The genes are classified into two groups of cancer factors and normal using a threshold value. The sklearn.metrics module has been used to fine-tuning the threshold value. Then f-measure and number of recognized CDGs by other methods are compared. The results of the proposed algorithm have been compared with the results of 18 computational and networking models in [24] to show the accurate performance of the proposed improved algorithm.

Figure 3 shows the results related to colon cancer. Two methods of comdp and memo have been removed from the diagram because of their zero values. As shown, KatzDriver has recognized 173 drivers for the colon cancer network, that it has the best performance among other network-based and computational (after ipac) methods. Moreover, it has the best performance among all the network-based and computational approaches by reaching f-measure=0.242. Figure 4 illustrates the comparison of KatzDriver with other methods for lung cancer network. We remove MDPfinder and memo methods because of their zero values. KatzDriver has the best performance among all the networking and computational methods similar to iMaxDriver-W by obtaining f-measure=0.242. Moreover, it has the first rank among the other 18 methods in this network by recognizing 155 drivers. The comparison of the proposed algorithm with other methods for the breast cancer network has been shown in figure 5. MDPfinder and comdp methods have been removed from the diagram due to their zero values. In this network, KatzDriver reaches f-measure=0.242 that has the best performance among other networking and computational methods. Moreover, it has recognized 152 drivers that has the best result among the network-based and computational methods after ipac.

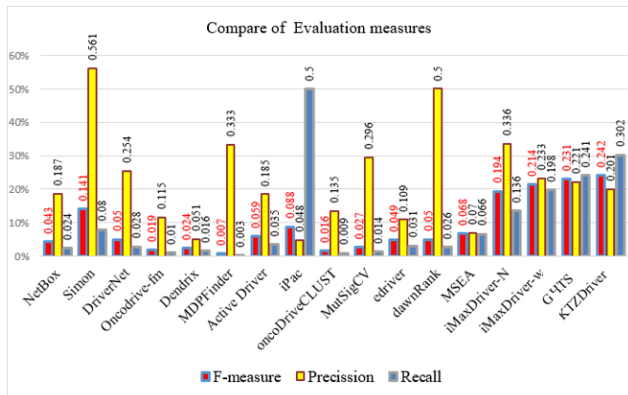
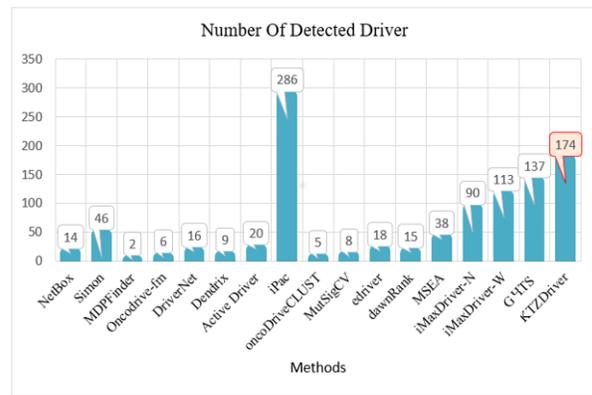

Figure 3: Results of KatzDriver with With 18 other methods for COAD network using CGC datasets.

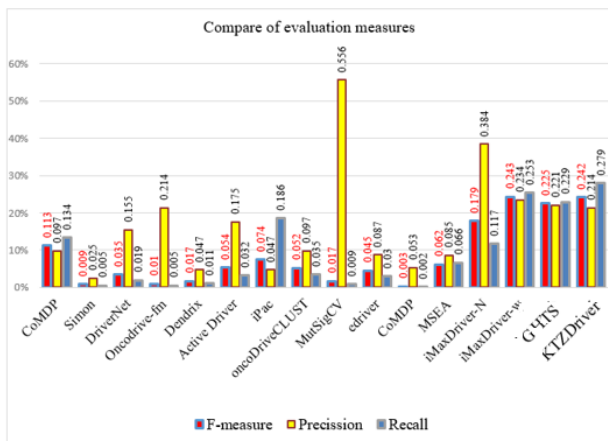
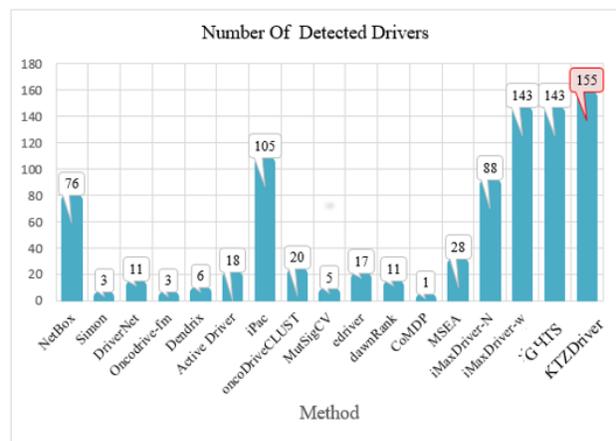

Figure 4: Results of KatzDriver with With 18 other methods for LUSC network using CGC datasets.

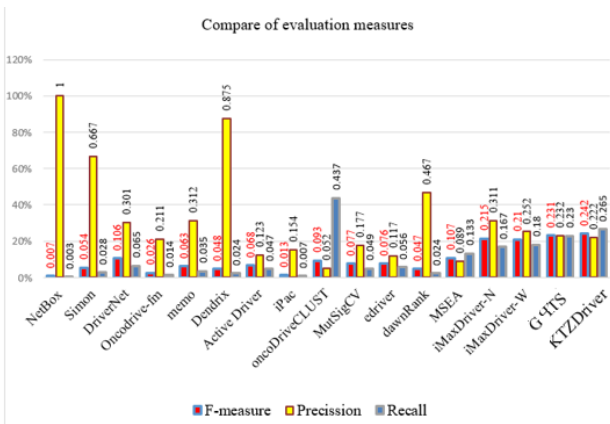
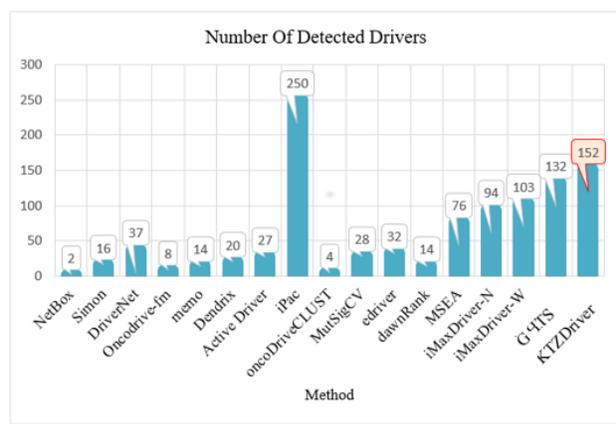

Figure 5: Results of KatzDriver with With 18 other methods for BRCA network using CGC datasets.

Venn diagrams of the mentioned cancer networks have been shown in figure 9 for more comparison of the recognized CDGs by KatzDriver and other methods. KatzDriver can recognize 34% of all the recognized genes by all the computational methods for colon cancer. Moreover, it has recognized 55 genes that are not detected by any of the computational techniques. So, it can be a proper complementary method for computational methods. Also, KatzDriver has recognized 108 drivers for lung cancer that none of the computational methods can recognize that it is better than the network-based techniques of iMaxDriver-N, iMaxDrver-W, and GHTS. In breast cancer, KatzDriver recognizes 79 of the drivers that they cannot be detected by any of the computational methods. Besides, it has detected 14, 8, and 13 unique genes in colon, lung, and breast cancer networks that none of the computational and network-based approaches have recognized. The binary matrix illustration of the genes was detected as driver by computational and networking methods using CGC database is shown in figure 6.

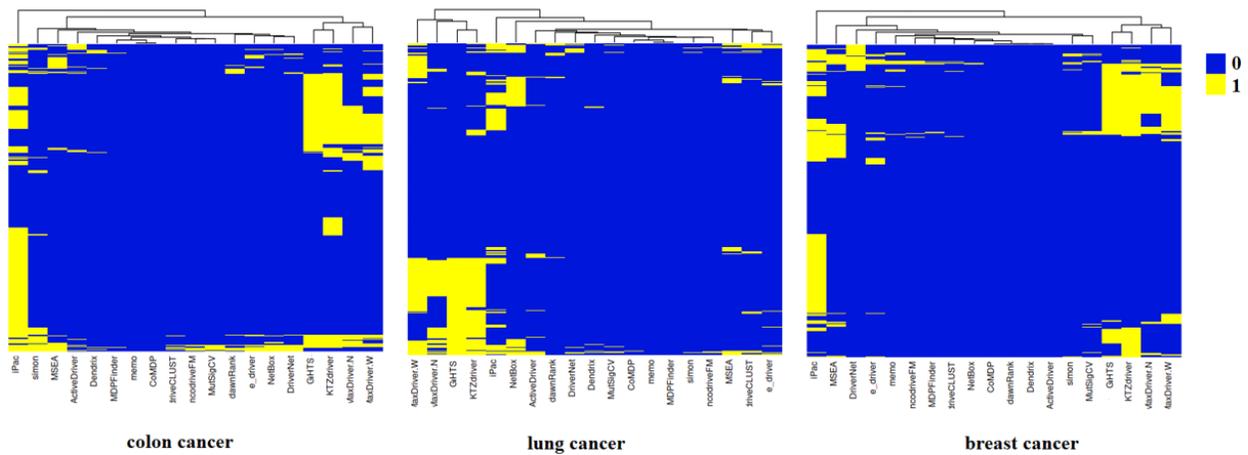

**Figure 6**: The binary matrix illustration of the genes was detected as driver by computational and networking methods using CGC.

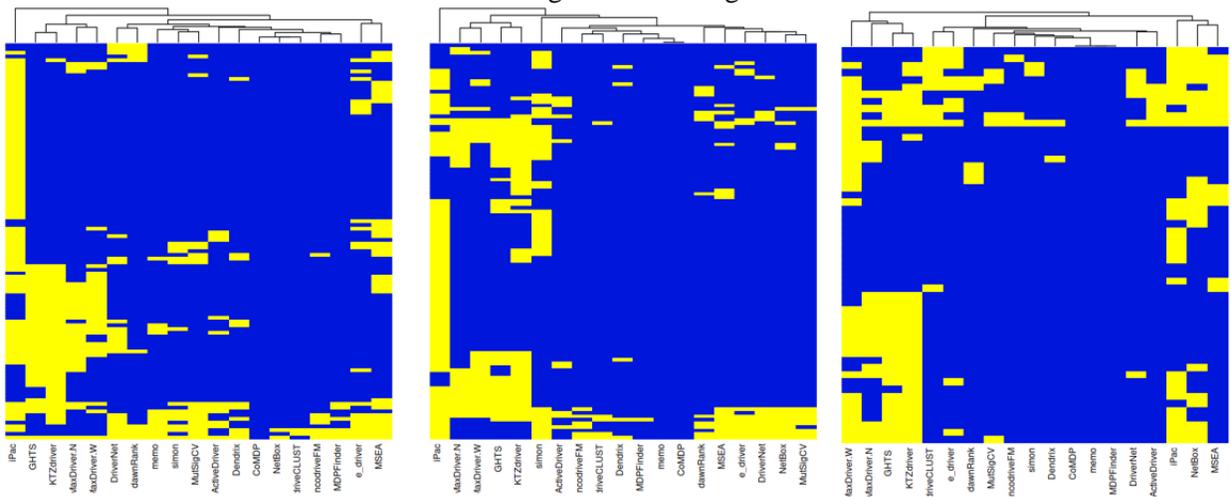

Figure 7: The binary matrix illustration of the genes was detected as driver by computational and networking methods using Mut-driver set.

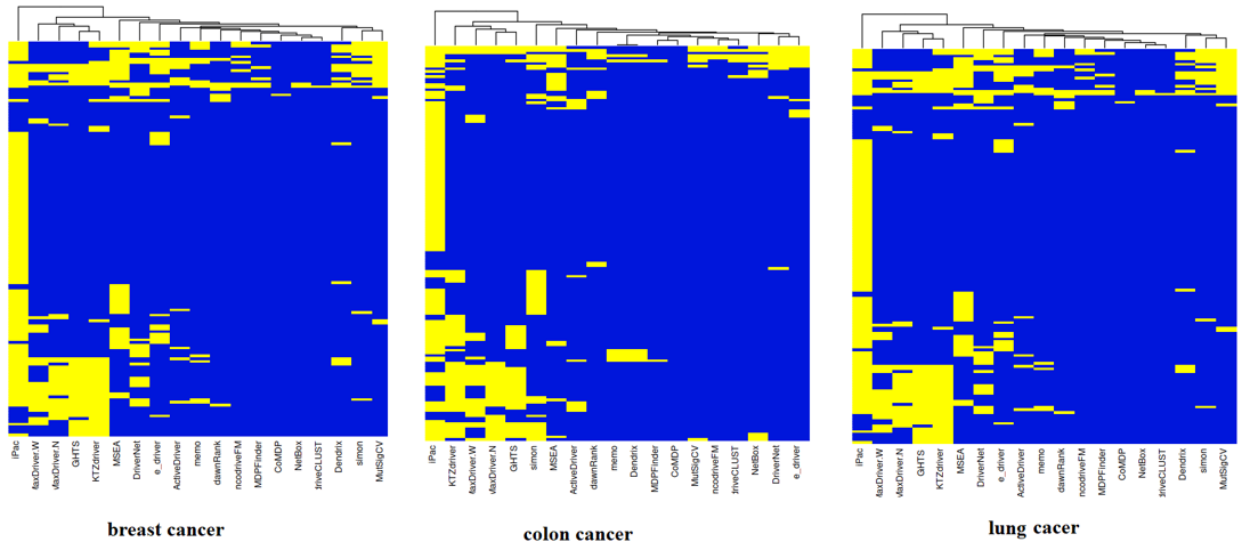

**breast cancer**   **colon cancer**   **lung cacer**

Figure 8: The binary matrix illustration of the genes was detected as driver by computational and networking methods using MSKCC set.

In addition to comparing the proposed method with the other methods using the CGC database in terms of the number of detected drivers and the fraction of detected drivers, for more validations we performed this comparison using the MSKCC dataset and the list of validated CDGs in Mut-driver and HiConf, as shown in table 2. For each of the validation datasets and each of the cancer type, top two approaches with the best detection results is shown as bold. In all of the cancer types, the KatzDriver is one of the top two approaches with the best results. In addition, the KatzDriver in all of the cancer types is better that the others network-based methods. It should be noted that, most prediction methods are able to detect a limit number of CDGs. Although some of these method like iPac predict many CDGs, but it has not an acceptable accuracy, as its f-measure and precision shown in figures 3-5.

**Table 1**: The data used to construct the network

| **GE dataset(GEO)** https://www.ncbi.nlm.nih.gov/geo/ | | | **Regulatory interactions (TRRUST)** https://www.grnpedia.org/trrust/ | | |
|---|---|---|---|---|---|
| GEO ID | Cancer type | Number of entries | Number on interactions | Number of genes | |
| | | | | # of TF genes | # of non-TF genes |
| GSE3268 | Lung cancer (LUSC) | 22,284 | 8,427 | 795 | 2,067 |
| GSE32323 | Colon cancer(COAD) | 54,675 | | | |
| GSE15852 | Breast cancer (BRCA) | 19,813 | | | |

**Table 2**: The comparison result of KatzDriver with other methods useing CGC, MSKCC,HiConf and Mut-Driver datasets.

| Validation Dataset | | Method Name | Netbox | Simon | Driveret | OncodriveFM | memo | Dendrix | MDPFinder | ActiveDriver | iPac | oncodriveCLUST | MutSigCV | e_driver | dawnRank | CoMDP | MSEA | iMaxDriver-N | iMaxDriver-W | GHTS | KatzDriver |
|---|---|---|---|---|---|---|---|---|---|---|---|---|---|---|---|---|---|---|---|---|---|
| Mut-Driver | BRCA | # predicted drivers | 2 | 13 | 22 | 7 | 12 | 11 | 6 | 11 | 81 | 2 | 17 | 20 | 10 | 0 | 33 | 29 | 28 | 38 | 45 |
| | | % predicted drivers | 1.6 | 10.4 | 17.6 | 5.6 | 9.6 | 8.8 | 4.8 | 8.8 | 64.8 | 1.6 | 13.6 | 16 | 8 | 0 | 26.4 | 23.2 | 22.4 | 30.4 | 36 |
| | COAD | # predicted drivers | 13 | 45 | 13 | 6 | 0 | 5 | 1 | 10 | 85 | 5 | 9 | 13 | 12 | 0 | 18 | 31 | 30 | 36 | 51 |
| | | % predicted drivers | 10.4 | 36 | 10.4 | 4.8 | 0 | 4 | 0.8 | 8 | 68 | 4 | 7.2 | 10.4 | 9.6 | 0 | 14.4 | 24.8 | 24 | 28.8 | 40.8 |
| | LUSC | # predicted drivers | 28 | 4 | 6 | 3 | 0 | 2 | 7 | 40 | 12 | 5 | 5 | 11 | 9 | 0 | 19 | 30 | 46 | 39 | 47 |
| | | % predicted drivers | 22.4 | 3.2 | 4.8 | 2.4 | 0 | 1.6 | 0 | 5.6 | 32 | 9.6 | 4 | 8.8 | 7.2 | 0 | 15.2 | 24 | 36.8 | 31.2 | 37.6 |
| HiConf | BRCA | # predicted drivers | 2 | 8 | 22 | 6 | 10 | 9 | 5 | 8 | 63 | 2 | 11 | 16 | 11 | 0 | 27 | 27 | 27 | 34 | 43 |
| | | % predicted drivers | 0.02 | 8.08 | 22.22 | 6.06 | 10.10 | 9.09 | 5.05 | 8.08 | 63.64 | 2.02 | 11.11 | 16.16 | 11.11 | 0 | 27.27 | 27.27 | 27.27 | 34.34 | 44.44 |
| | COAD | # predicted drivers | 11 | 37 | 12 | 6 | 0 | 5 | 1 | 10 | 63 | 4 | 8 | 11 | 12 | 0 | 20 | 28 | 29 | 34 | 50 |
| | | % predicted drivers | 11.11 | 37.37 | 12.12 | 6.06 | 0 | 5.05 | 1.01 | 10.1 | 63.64 | 4.04 | 8.08 | 11.11 | 12.12 | 0 | 20.20 | 28.28 | 29.29 | 34.34 | 50.51 |
| | LUSC | # predicted drivers | 21 | 3 | 7 | 3 | 0 | 1 | 0 | 9 | 32 | 11 | 5 | 9 | 10 | 0 | 17 | 26 | 43 | 36 | 45 |
| | | % predicted drivers | 21.21 | 3.03 | 7.07 | 3.03 | 0 | 1.01 | 0 | 9.09 | 32.32 | 11.11 | 5.05 | 9.09 | 10.10 | 0 | 17.17 | 26.26 | 43.43 | 36.36 | 45.45 |
| MSKCC | BRCA | # predicted drivers | 2 | 16 | 26 | 7 | 12 | 15 | 5 | 12 | 117 | 1 | 20 | 26 | 8 | 1 | 38 | 30 | 28 | 36 | 44 |
| | | % predicted drivers | 1.02 | 8.12 | 13.20 | 3.55 | 6.09 | 7.61 | 2.54 | 6.09 | 59.39 | 0.51 | 10.15 | 13.20 | 4.06 | 0.51 | 19.29 | 15.23 | 14.21 | 18.27 | 22.34 |
| | COAD | # predicted drivers | 16 | 51 | 17 | 6 | 0 | 10 | 2 | 18 | 208 | 6 | 9 | 19 | 22 | 0 | 36 | 61 | 69 | 68 | 96 |
| | | % predicted drivers | 2.84 | 8.98 | 2.60 | 0.95 | 0 | 1.65 | 0.47 | 2.84 | 27.66 | 1.18 | 1.65 | 2.84 | 2.36 | 0 | 4.96 | 7.09 | 6.62 | 8.27 | 11.82 |
| | LUSC | # predicted drivers | 60 | 4 | 15 | 3 | 0 | 4 | 0 | 11 | 82 | 17 | 6 | 17 | 18 | 0 | 27 | 57 | 97 | 72 | 87 |
| | | % predicted drivers | 12.09 | 1.01 | 2.77 | 0.76 | 0 | 1.01 | 0 | 1.51 | 13.10 | 3.27 | 1.51 | 4.03 | 1.76 | 0 | 4.79 | 7.30 | 11.84 | 9.32 | 11.34 |
| CGC | BRCA | # predicted drivers | 2 | 16 | 37 | 8 | 14 | 20 | 0 | 27 | 250 | 4 | 28 | 32 | 14 | 0 | 76 | 94 | 103 | 132 | 152 |
| | | % predicted drivers | 0.34 | 2.79 | 6.46 | 1.39 | 2.44 | 3.49 | 0 | 4.72 | 43.70 | 0.69 | 4.89 | 5.59 | 2.44 | 0 | 13.28 | 16.43 | 18 | 23.07 | 26.57 |
| | COAD | # predicted drivers | 14 | 46 | 2 | 6 | 0 | 16 | 9 | 20 | 286 | 5 | 8 | 18 | 15 | 0 | 38 | 90 | 113 | 137 | 174 |
| | | % predicted drivers | 2.44 | 8.04 | 0.34 | 1.04 | 0 | 2.79 | 1.57 | 3.49 | 50 | 0.87 | 1.39 | 3.14 | 2.62 | 0 | 6.64 | 15.73 | 19.75 | 23.95 | 30.41 |
| | LUSC | # predicted drivers | 76 | 3 | 11 | 3 | 6 | 18 | 0 | 105 | 20 | 5 | 17 | 11 | 1 | 0 | 28 | 88 | 143 | 143 | 155 |
| | | % predicted drivers | 13.42 | 0.53 | 1.94 | 0.53 | 1.06 | 3.18 | 0 | 18.55 | 3.53 | 0.88 | 3 | 1.94 | 0.17 | 0 | 4.94 | 15.54 | 25.26 | 25.26 | 27.38 |

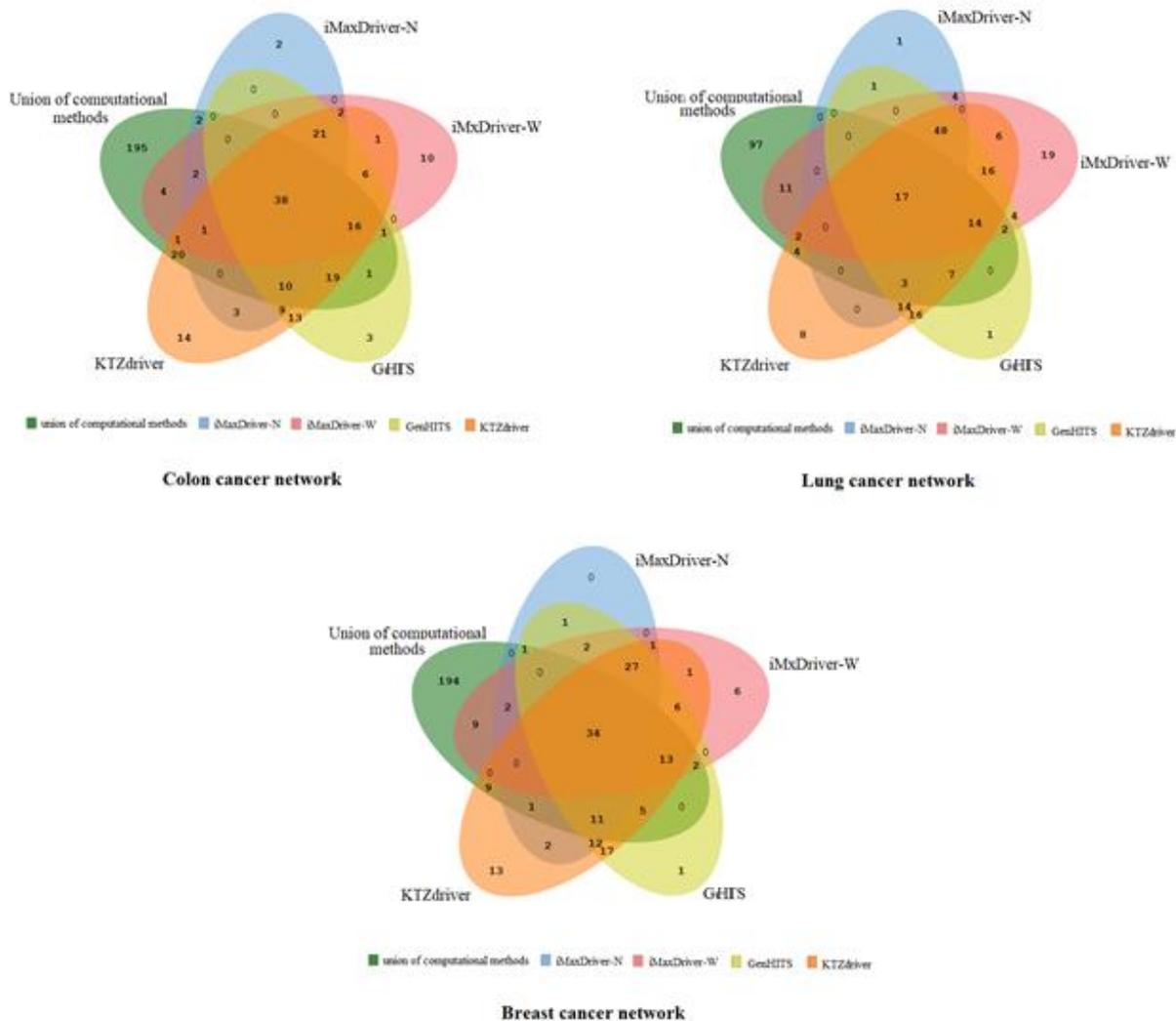

Figure 9: Venn diagrams of CDGs predicted by KatzDriver and other methods using CGC datasets.

In addition, we compared the KatzDriver's with iMaxDriver-N, iMaxDrver-W, and GHTS network-based methods based on the overlap rate of the detected CDGs. Compare results shown in figure 10. It recognizes 94%, 76%, and 97% of the recognized drivers for colon cancer by iMaxDriver-N, iMaxDrver-W, and GHTS, respectively. Moreover, it has detected 34 unique genes that cannot be recognized by three mentioned network-based methods. Also, KatzDriver has detected 94%, 73%, and 96% of the recognized drivers for lung cancer by iMaxDriver-N, iMaxDrver-W, and GHTS, respectively. For breast cancer, it has recognized 94%, 80%, and 94% of the recognized drivers by iMaxDriver-N, iMaxDrver-W, and GHTS, respectively. It has detected 22 unique genes that are not recognizable by the other three network-based methods. So it reaches the first rank among the network-based techniques.

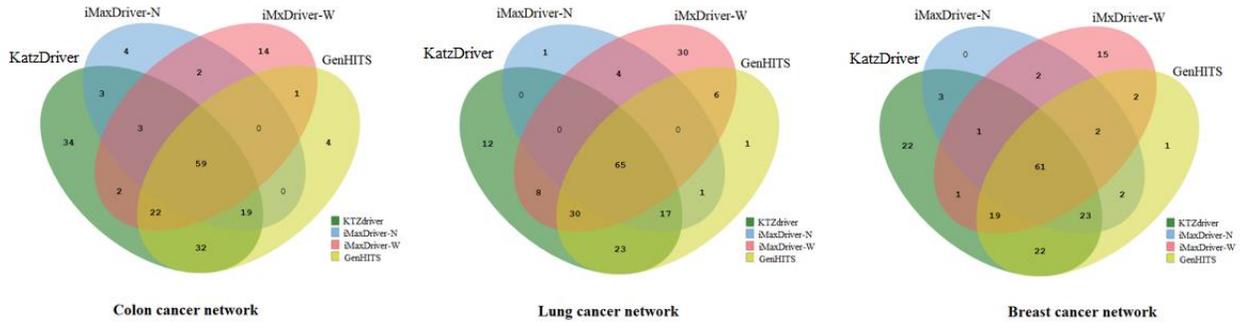

Figure 10: Venn diagrams of CDGs predicted by KatzDriver and other network based methods using CGC datasets.

## Conclusion

In this paper, we have proposed a network-based approach, namely KatzDriver, to recognize CDGs in the gene regulatory networks. It ranks the genes based on their relative importance for information propagation and reception. So, the genes with the highest effects on information propagation and reception, are the most effective ones in anomaly propagation in gene network and result in cancer propagation. In the preprocessing stage, the regulatory interactions are filtered, then the genes and the interactions are weighted based on biological data. The algorithm applied to three cancer network and the nodes are ranked. The genes with higher rank are more important and are classified as cancer genes. The results compared with 18 computational and networking methods. The most advantages of the proposed algorithm are independence on genomic data and mutation, higher precision among all the network and computational methods, recognition of more CDGs, and no need for heavy calculations like iMaxDriver methods. In addition to the recognition of most of the detected genes by other network-based methods, KatzDriver has the best performance among other network-based methods by detection of some unique genes. Moreover, compare with other network-based methods, it can recognize more genes that are not predictable by computational approaches. Hence, it can be used as the best network-based method, along with computational prediction tools. As future work, to improve the performance of the proposed method can improve the regulatory interaction network to have more regulatory interactions. The ability to add other molecule elements of the cell regulatory network like miRNA to the interaction network is different based on the kind of regulatory interaction with TFs and miRNAs. The proposed approach can be used in other cell networks like gene-disease to find molecular treatments.

# Bibliography

[1] Tokheim, C. J., Papadopoulos, N., Kinzler, K. W., Vogelstein, B., & Karchin, R. (2016). Evaluating the evaluation of cancer driver genes. Proceedings of the National Academy of Sciences, 113(50), 14330-14335.

[2] Rao, V. S., Srinivas, K., Sujini, G. N., & Kumar, G. N. (2014). Protein-protein interaction detection: methods and analysis. International journal of proteomics, 2014.

[3] MacNeil, L. T., & Walhout, A. J. (2011). Gene regulatory networks and the role of robustness and stochasticity in the control of gene expression. Genome research, 21(5), 645-657.

[4] Katz, L. (1953). A new status index derived from sociometric analysis. Psychometrika, 18(1), 39-43.

[5] Mark E. J. Newman: Networks: An Introduction. Oxford University Press, USA, 2010, p. 720.

[6] Han, H., Cho, J.W., Lee, S., Yun, A., Kim, H., Bae, D., Yang, S., Kim, C.Y., Lee, M., Kim, E. and Lee, S., 2017. TRRUST v2: an expanded reference database of human and mouse transcriptional regulatory interactions. Nucleic acids research, 46(D1), pp.D380-D386.

[7] Tusher, V. G., Tibshirani, R., & Chu, G. (2001). Significance analysis of microarrays applied to the ionizing radiation response. Proceedings of the National Academy of Sciences, 98(9), 5116-5121.

[8] Barrett, T., Wilhite, S.E., Ledoux, P., Evangelista, C., Kim, I.F., Tomashevsky, M., Marshall, K.A., et al. (2012), "NCBI GEO: archive for functional genomics data sets—update", Nucleic Acids Research, Vol. 41 No. D1, pp. D991–D995.

[9] Futreal, P. A., Coin, L., Marshall, M., Down, T., Hubbard, T., Wooster, R., & Stratton, M. R. (2004). A census of human cancer genes. Nature reviews cancer, 4(3), 177-183.